# Security Analysis of Smart Contract Migration from Ethereum to Arbitrum

**Xueyan Tang[1,2], Lingzhi Shi[1]**

[1]Salus Security, Beijing, 100020, China
[2]corresponding author, 777728@gmail.com

## ABSTRACT

When migrating smart contracts from one blockchain platform to another, there are potential security risks. This is because different blockchain platforms have different environments and characteristics for executing smart contracts. The focus of this paper is to study the security risks associated with the migration of smart contracts from Ethereum to Arbitrum. We collected relevant data and analyzed smart contract migration cases to explore the differences between Ethereum and Arbitrum in areas such as Arbitrum cross-chain messaging, block properties, contract address alias, and gas fees. From the 36 types of smart contract migration cases we identified, we selected 4 typical types of cases and summarized their security risks. The research shows that smart contracts deployed on Ethereum may face certain potential security risks during migration to Arbitrum, mainly due to issues inherent in public blockchain characteristics, such as outdated off-chain data obtained by the inactive sequencer, logic errors based on time, the permission check failed, Denial of Service(DOS) attacks. To mitigate these security risks, we proposed avoidance methods and provided considerations for users and developers to ensure a secure migration process. It's worth noting that this study is the first to conduct an in-depth analysis of the secure migration of smart contracts from Ethereum to Arbitrum.

## Introduction

Smart contracts are automated execution programs running on blockchain platforms. With the continuous development and application promotion of blockchain technology, smart contract migration has become a topic of considerable interest[1]. Smart contract migration refers to the process of transferring a contract from one blockchain platform to another. It holds significant value in the blockchain realm[2], enabling cross-chain interoperability among various blockchain platforms and facilitating the free flow of data and assets across different platforms[3]. Through migration, projects can leverage the technology of new blockchain platforms to enhance the performance, security, and scalability of their contracts. We will conduct research focusing on the smart contract migrating from Ethereum[4] to Arbitrum[5].

However, smart contract cross-platform migration faces security risks. Different blockchain platforms may adopt varying protocols, standards, smart contract languages, and virtual machines. This could result in a smart contract without any apparent flaws, when deployed on one blockchain platform, encountering security issues when executed on another platform. Therefore, to better mitigate potential security risks and ensure smooth contract migration[6], it's necessary to thoroughly assess the characteristics of the destination chain and the underlying differences between the source and destination chains prior to the migration[7]. The primary objective of our study is to help developers understand the underlying differences between Ethereum[4] and Arbitrum[5], thereby enhancing the safety of smart contract migration.

The contributions of this study are as follows:

1. We identify security risks related to the runtime environment of smart contracts, highlight the importance of understanding the underlying differences between the source and target blockchains during smart contract migration for vulnerability identification.

2. Through a multifaceted analysis of differences, we delved into the key distinctions between Ethereum and Arbitrum. Our research revealed several unique aspects, including Arbitrum Cross-Chain Messaging, Block Properties, Contract Address Alias, and Gas Fees.

3. Through multiple case studies, we summarize the potential risks that may arise during the migration of smart contracts from Ethereum to Arbitrum. These include outdated off-chain data obtained by the inactive sequencer, logic errors based on time, the permission check failed, DOS attack. 4. To address these issues, we provide relevant mitigation strategies, which are valuable
for users and developers with smart contract migration needs. It is worth mentioning that this paper is the first in-depth research on the secure migration of smart contracts from Ethereum to Arbitrum.

## Background

### Ethereum

Ethereum[4] is a platform that integrates distributed ledger technology, cryptographic techniques, consensus algorithms, and smart contracts. Its design goal is to achieve decentralization, security, transparency, and programmability. As of October 2023, Ethereum is one of the most widely used blockchain platforms, with a market value of approximately $167.46 billion[8]. It provides the foundation and innovative opportunities for smart contracts, decentralized finance, crowdfunding, and token issuance, and has a strong community and developer ecosystem.

Smart Contracts[9]: Smart contracts are programming protocols used to automatically execute and verify contract terms. They are mainly used for executing and managing transactions and the functionalities of digital assets. This concept was first introduced to the blockchain field by Ethereum. Ethereum smart contracts are typically developed using the Solidity programming language[10].

Solidity Programming Language[10]: Solidity is a Turing-complete programming language that allows developers to implement various complex functions and logic using powerful features such as loop control structures, conditional statements, and recursive calls. In addition, Solidity supports advanced language features such as function calls, modifiers, overloading, events, inheritance, and libraries, greatly enhancing the programmability and flexibility of contracts.

Smart Contract Deployment: Smart contracts are deployed on blockchain platforms such as Ethereum. After deployment, a smart contract is assigned a unique contract address, which is used to identify and locate instances of the contract on the blockchain. Successfully deployed smart contract code is converted into bytecode that can be executed on the EVM[11].

EVM[11]: The EVM is a stack-based virtual machine with its own instruction set and memory model. It is a fully isolated runtime environment for executing smart contracts and maintaining the consistency of the Ethereum platform. Its design goal is to provide a secure and reliable environment. The EVM uses a gas-based mechanism to limit the computational and storage resource consumption of contracts.

Gas Fee[12]: Gas fee is used to measure the computational complexity and resource consumption of contract execution. Each operation and instruction has a predefined gas cost, and the actual consumption is billed accordingly during execution. The gas mechanism ensures that the execution of smart contracts does not consume resources without limits, preventing issues such as malicious code or infinite loops. Paying sufficient gas fees ensures the smooth execution of smart contracts and correct result returns.

### Arbitrum

Arbitrum, developed by Offchain Labs, is a novel Ethereum L2 scaling solution that has gained significant attention and popularity on Ethereum community. Arbitrum supports fast smart contract transactions while reducing transaction costs. Arbitrum is highly compatible with the Ethereum Virtual Machine (EVM)[13]. Developers can write smart contracts using Ethereum tools like Solidity and easily migrate them to Arbitrum. Therefore, Arbitrum is also a popular choice for smart contract migration.

As of August 31, 2023, Arbitrum has captured 55.95% of the L2 market share[5]. There are two Arbitrum chains on Ethereum mainnet: Arbitrum One, an Arbitrum Rollup chain, and Nova, an AnyTrust chain. The data availability for Arbitrum One is on-chain (Ethereum mainnet), while the data availability for Nova is off-chain (Data Availability Committee).

The Arbitrum One Portal stores off-chain execution data on Ethereum mainnet. As of October 2023, a total of 576 projects have been deployed on Arbitrum One[14], such as Curve and OpenSea. As of October 2023, according to official statistics from Coingecko, the Total Value Locked (TVL) in Arbitrum One is over $1.7 billion, making it the largest market share among all L2 chains today[15]. TVL is a measure of the total value of assets locked by users on decentralized finance (DeFi) platforms or applications. As shown in Figure 1, Arbitrum has the highest TVL share in L2.

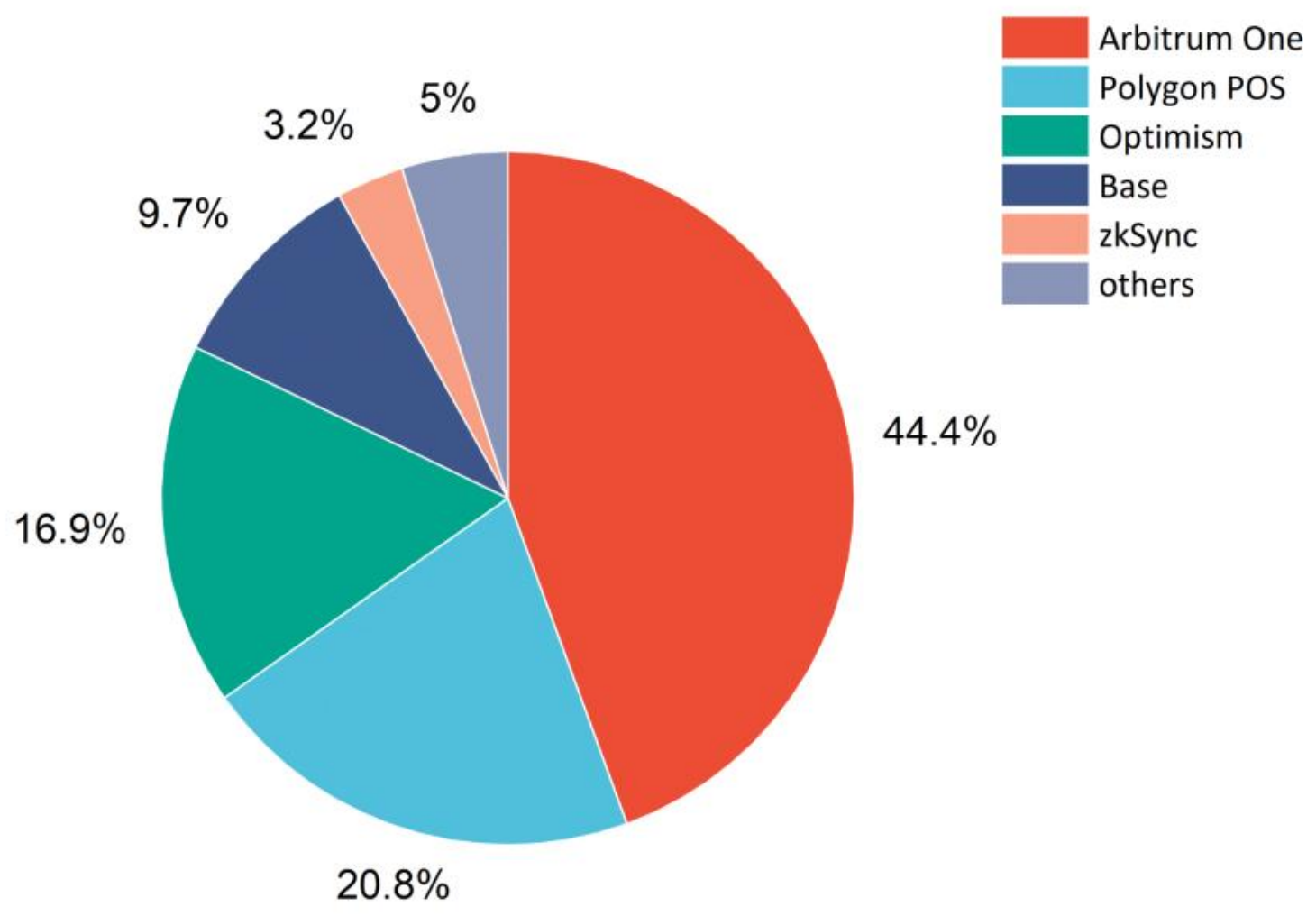

**Figure 1.** Distribution of TVL in L2. Arbitrum One has the largest share, followed by Polygon Pos and Optimism.

## The Migration of Smart Contracts from L1 to L2

In this section, we primarily discuss the process of migrating smart contracts from Ethereum to L2, as well as the key features of these L2 related to smart contracts, such as the development language and runtime environment for smart contracts.

### *Smart Contract Migration Process*

Smart contract migration mainly consists of two steps: recovering the data to be migrated and writing this data into a new contract deployed on the target blockchain. Figure 2 is Smart Contract Migration Process Diagram.

1. Data Recovery: The first step in smart contract migration is data recovery. In this step, we need to read data from specific blocks on the source blockchain for migration and use corresponding methods to recover the data based on the contract's data structure. For example, by calling the appropriate getter() function to retrieve public variables from the contract, relying on events, or using the getStorageAt() function to calculate the memory offset of private variables and retrieve their values from contract storage. However, please note that in order to increase the transparency of the migration and prevent attackers from exploiting users who are unaware of the migration, it is recommended to pause the contract's operation during the data recovery process.

2. Writing Data to New Contract: During smart contract migration, once the data to be migrated has been collected, the next step is to deploy and initialize the new contract on the target blockchain for migration. For simple variables, their values can be set through the contract's constructor() function. If the amount of data being migrated is large, the migration needs to be split into multiple transactions. For example, for ERC20 token contracts, the contract needs to be deployed on the target blockchain and the initial parameters set in the initial state. Then, users initiate transfer transactions in the old contract, and the new contract receives and records these transfer operations, moving the token balances from the old contract to the new one.

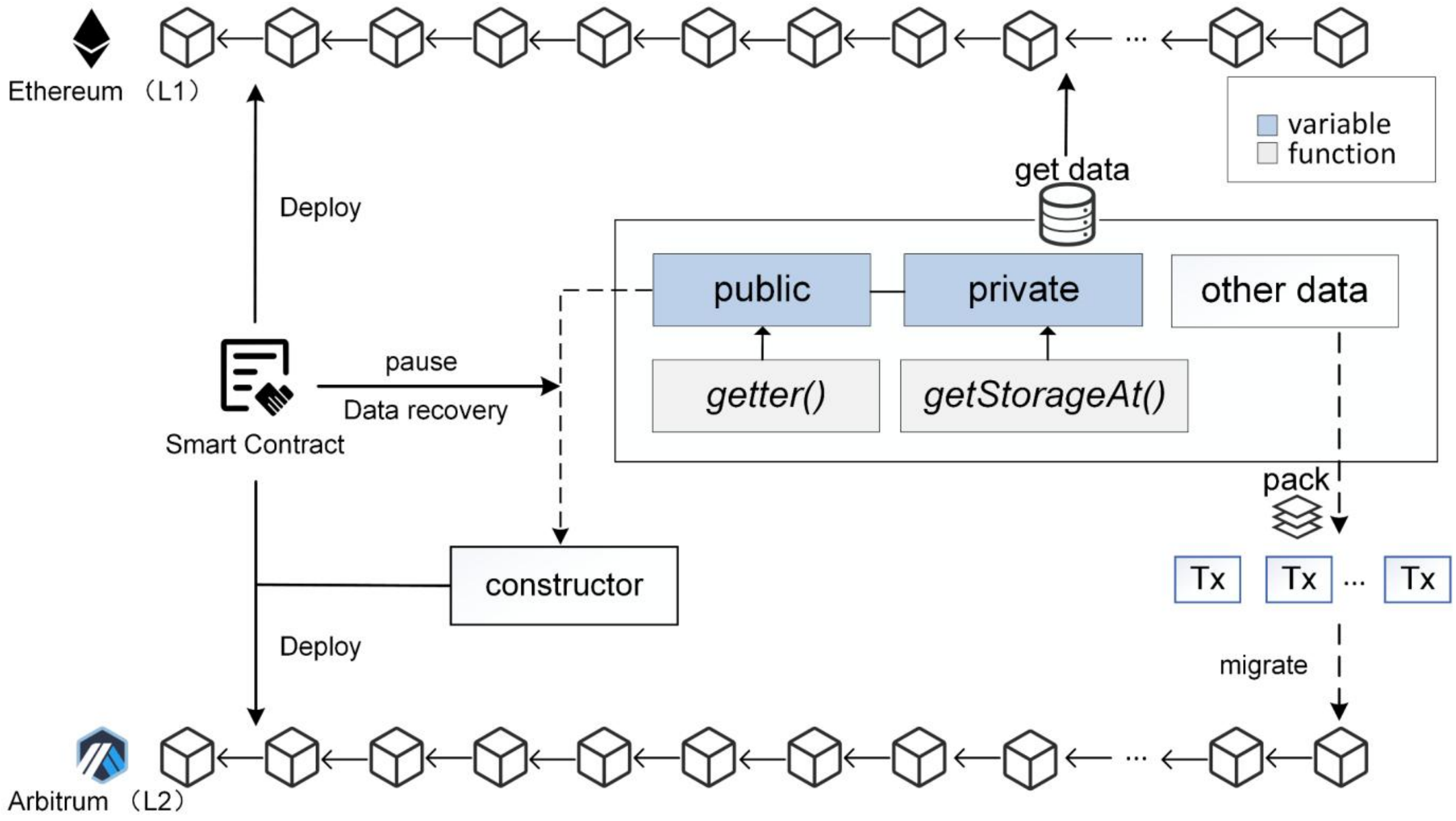

**Figure 2.** Smart Contract Migration Process Diagram. Migrating smart contracts deployed on Ethereum to Arbitrum, where Ethereum is the source blockchain for migration and Arbitrum is the target blockchain.

***Differences Between L1 and L2***

This paper mainly focuses on the case study of migrating smart contracts deployed on Ethereum to Arbitrum, revealing the security risks caused by migration. However, there are many smart contract migration solutions that are not limited to Arbitrum. L2 chains such as Polygon and StarkNet are also popular migration target blockchains. We explain the differences between L1 and L2 from the following two aspects:

1. Smart Contract Development Language: There are many programming languages for smart contract development, including Solidity[10], Vyper[16], Cairo[17], and more. Solidity is widely used for smart contract development on Ethereum and has a large developer community. Vyper is also an officially supported language for Ethereum. Most L2 blockchains also support Solidity, such as Arbitrum, Optimism[18]. Some L2 blockchains have their own smart contract development languages, such as Cairo for smart contracts on StarkNet[19]. Cairo is a low-level intermediate language used to describe the computation and state transition logic of StarkNet.

2. Smart Contract Runtime Environment: On Ethereum, the runtime environment for smart contracts based on Solidity and Vyper is the EVM. Many L2 solutions provide runtime environments for smart contracts that are EVM compatible, but the differences between them should not be overlooked. For example, Arbitrum's smart contracts[20] runtime environment simulates EVM execution through its custom Arbitrum Virtual Machine (AVM) or the WebAssembly-based ArbOS program. These execution environments aim to provide an execution environment that closely resembles the EVM. The operating environment for Polygon zkEVM[21] is the zkEVM developed by Polygon. Differences between Polygon zkEVM and EVM mainly include opcodes, precompiled contracts, and so on[22].

**Vulnerabilities in smart contracts**

Vulnerabilities in smart contracts refer to security weaknesses or flaws in the code of a smart contract, which can lead to unauthorized access, loss of funds, data tampering, or other adverse consequences. Some common smart contract vulnerabilities include weak random number generation, access control flaws, DOS attacks, and reentrancy.

Access Control Flaws occur when there are errors or inappropriate permission control mechanisms in a smart contract, allowing unauthorized or malicious users to execute operations or access data they shouldn't have access to.

DOS Attacks occur when malicious users exploit flaws or vulnerabilities in a contract to consume a large amount of resources or perform time-consuming operations, causing the contract to malfunction or block other users' operations.

Reentrancy refers to a vulnerability in which a smart contract fails to properly handle the invocation of an external contract, resulting in a reentrant call to a function within the current contract during the execution of the external contract. This type of vulnerability can lead to incorrect data state or financial loss within the contract.

Weak Randomness refers to the use of insecure or unreliable methods for generating random numbers in a smart contract, allowing attackers to predict or manipulate the outcome of random numbers and gain an unfair advantage.

The security of smart contracts extends beyond common vulnerabilities and also includes contract migration. Due to differences in implementation details, security features, and programming models between different platforms and runtime environments, these differences could potentially lead to security risks. To ensure the safe execution of smart contracts when migrating from Ethereum to Arbitrum, we investigate the underlying differences between these two environments, the features of Arbitrum, and the vulnerabilities in smart contracts caused by these differences. While these vulnerabilities also fall under common smart contract vulnerabilities, they cannot be easily identified without understanding the characteristics of the smart contract's runtime environment.

## Related work

Smart contract migration is the process of transferring deployed smart contracts from one blockchain platform to another. Secure smart contract migration is an area of great interest. In fact, there have been some real-life cases that demonstrate the feasibility of smart contract migration. For example, Tank NFT (ERC721 Token Smart Contract) successfully migrated from Ethereum to Binance Smart Chain (BSC)[23]. BasketCoin also successfully migrated from Ethereum to BSC[24]. The "The Sandbox" game project also migrated its smart contracts to the Polygon platform[25]. Blox NFT migrated from Ethereum to Arbitrum[26]. These cases demonstrate that smart contract migration is feasible and further highlight the importance and value of smart contract migration to reputable and promising blockchain platforms.

These cases indicate that smart contract migration is feasible, but special attention must be paid to security issues during the migration process. For example, the migration of the Uniswap governance system resulted in the inability to execute governance proposals. Due to poor communication among developers and misconfiguration of the system, the Uniswap governance system has encountered serious issues during the migration to Arbitrum, resulting in the inability to execute governance proposals on Arbitrum[27].

Furthermore, academic research provides us with valuable guidance. M. Westerkamp proposed a toolbox that enables the migration of smart contracts between EVM-compatible blockchains, using the migration of an ERC20 token contract as an example to demonstrate the effectiveness of the toolbox[28]. However, this research focuses on the verifiability of the code and state of the smart contract to be migrated, as well as the mechanism of smart contract migration using their proposed toolbox. They do not discuss the security issues of migration from the perspective of the smart contract. In contrast, our work focuses on security risks of the smart contract to be migrated. K. Shudo et al. emphasize the importance of application portability. They design middleware to facilitate the migration of applications between different blockchains to address the issue of mismatched incentives in public blockchains[29]. However, they also do not touch upon the portability and security of smart contract code and data.

## Methodology

We have employed a mixed approach to study the security migration of smart contracts, which includes data collection and organization, case study, and expert involvement.

### Data Collection

There are three main channels for data collection: relevant literature[30], official online resources[31], and smart contract source code cases[32]. These resources provide the latest research findings and industry dynamics, enabling us to understand cutting-edge knowledge in the field of smart contract security migration. We also categorized all the collected materials, creating directories based on different themes and keywords. Relevant official online resources and smart contract source codes were classified into corresponding categories.

- Collection of Relevant Literature: During the process of collecting relevant literature, we used keywords such as "Smart Contract", "Migration" and "Security" to conduct literature searches on Google Scholar. We found that the number of relevant literature in the field of smart contract security migration is limited, which may be due to its relatively new and technologically complex nature. There are two papers[28, 29] that focus on discussing migration mechanisms by designing toolkits and middleware to facilitate the migration of smart contracts or applications. However, they do not provide explanations and verifications regarding the security of migrating smart contracts and applications.

- Collection of Official Online Resources: Although the amount of literature specifically focused on smart contract migration is limited, we made efforts to gather a large amount of official online resources to supplement the literature. These resources have been extremely helpful for our research. Since our focus is on smart contract migration from Ethereum to Arbitrum, we primarily relied on the official websites of Ethereum and Arbitrum for data collection[33-39]. In Particular, we found numerous relevant online resources on Arbitrum website, highlighting the differences between Arbitrum and Ethereum. Additionally, we collected data from the official websites of OpenZeppelin[40], Chainlink[41], and Solidity[10].

- Collection of Smart Contract Source Code: We collected cases of smart contract vulnerabilities from audit reports found on the official websites, including Code4rena[42], Cyfrin[43], Halborn[44], OpenZeppelin[40] and so on, encompassing a total of 36 types of smart contracts. We also referred to swcregistry (classification of smart contract vulnerability types), Ethereum Smart Contract Best Practices–Attacks, and the UNISWAP project. Table 1 displays some of our data sources.

| Data Platforms | Reference |
|---|---|
| Google Scholar | [28] [29] |
| Online Resources | [45] [33] [34] [35] [36] [37] [38] [39] [46] [47] |
| Smart Contract Vulnerability Cases | [48] [49] [50] [51] [52] [53] [54] [55] |

**Table 1.** Source of Data

**Data Organization**

Our team organized and categorized the collected materials and conducted case analysis. Based on the relevance of the content, the collected official online resources and smart contract source code will be categorized into 4 categories: Arbitrum cross chain messaging, block properties, address alias and gas fees. This classification method is mainly based on the supporting materials of the Offchain Labs team[33-39], who are the core leaders and development team of Arbitrum. It's worth noting that the description of the main differences between Arbitrum and Ethereum on the official Arbitrum website also primarily focuses on these 4 categories.

**Case Study**

Based on the organized materials, our team conducted an analysis of smart contract source code cases during the migration process from Ethereum to Arbitrum. Each member of our team has over two years of experience in smart contract development and auditing. They audited the collected source code cases and analyzed their security risks.

During our case analysis, we underwent the following four steps:

1. Analyze the underlying differences and characteristics between Ethereum and Arbitrum.

2. Based on the differences between Ethereum and Arbitrum's foundations, examine the smart contract source code to understand the contract's inheritance relationships, function call relationships, and various variables.

3. Combine automated vulnerability detection tools with manual auditing, and conduct group discussions on the audit results.

4. Synthesize case studies and foundational differences to summarize the potential security risks in smart contracts migrating from Ethereum to Arbitrum.

We organized a total of 36 types of related cases. These cases include smart contract source code with vulnerabilities, contract functionality, the causes of vulnerabilities, and Avoidance methods provided by our auditing experts. From the 36 types of smart contract migration cases we collected, we selected four typical cases. Among these, there is one case related to Arbitrum Cross Chain Messaging, two cases concerning Block Properties, one case on Contract

Address Alias, and two cases about Gas Fees. These results are presented in the supplementary information.

### Expert participation

To ensure the accuracy and comprehensiveness of our information collection, analysis, and organization, we have invited researchers from Offchain Labs to participate in our research. In the initial stages of the research, the experts provided us with targeted reference materials, which allowed us to conduct our research with a clear purpose. As our research reached milestone stages, due to physical distance, we communicated with the experts through online meetings and used document collaboration to guide our work.

The expertise and experience of these experts have provided us with valuable insights and recommendations[56], helping us gain a deeper understanding of the challenges and solutions involved in securely migrating smart contracts.

## Findings

Based on the academic literature, official online resources, and relevant smart contract vulnerability cases[57] we have collected, and under the guidance of experts, we have made the following findings: Although the Ethereum mainnet and the L2 scaling solution Arbitrum are both EVM-compatible blockchain platforms, there are still some noteworthy differences between them. These underlying differences can have an impact on the execution of smart contracts and may lead to security issues. In this section, we will explain the underlying differences between Ethereum and Arbitrum, as well as the impact these differences will have on smart contract migration from four aspects: Arbitrum Cross-Chain Messaging, Block Properties, Contract Address Alias and Gas Fees.

### Arbitrum Cross-Chain Messaging

Arbitrum Cross-Chain Messaging is a mechanism for cross-chain interaction between Ethereum and Arbitrum. Through this mechanism, users can exchange messages and data between these two platforms to facilitate asset transfers, transaction execution, and other operations. Whether users are operating on Ethereum or Arbitrum, they can fully leverage the characteristics and ecosystems of each platform to achieve more efficient and secure cross-chain interactions.

Under the Arbitrum Cross-Chain Messaging mechanism, cross-chain message passing between Ethereum (referred to as L1) and Arbitrum (referred to as L2) is divided into two types: L2-to-L1 and L1-to-L2 transactions, as shown in Figure 3. Understanding these two transaction types and their mechanisms is crucial for ensuring smooth transaction execution and security. Different roles and contracts play important roles in these transaction types.

L2-to-L1 transactions refer to transfers or transaction operations from the Arbitrum to the Ethereum. In this transaction type, the Arbitrum sequencer plays a crucial role. The sequencer is a special node responsible for batching transactions on Arbitrum and submitting them to the Arbitrum on Ethereum. The role of the sequencer is to ensure that transactions on Arbitrum enter the Ethereum in the correct order, ensuring transaction correctness and security. Therefore, ensuring the smooth operation of the sequencer node is essential.

L1-to-L2 transactions are transfers or transaction operations from Ethereum to Arbitrum. This type of transaction is facilitated through the Bridge mechanism. Users send assets from Ethereum to a specific Ethereum account, and then move these assets to the corresponding account on Arbitrum through the bridge. In L1-to-L2 transactions, retryable tickets also play a crucial role. Retryable tickets are special contracts used to ensure that transfers or transaction operations on Ethereum can be successfully executed on Arbitrum. If a transaction fails on Arbitrum, users can use the retryable tickets to retry the execution of that transaction. The preset gas fees in the L1-to-L2 message call process determine whether the transaction can be successfully executed.

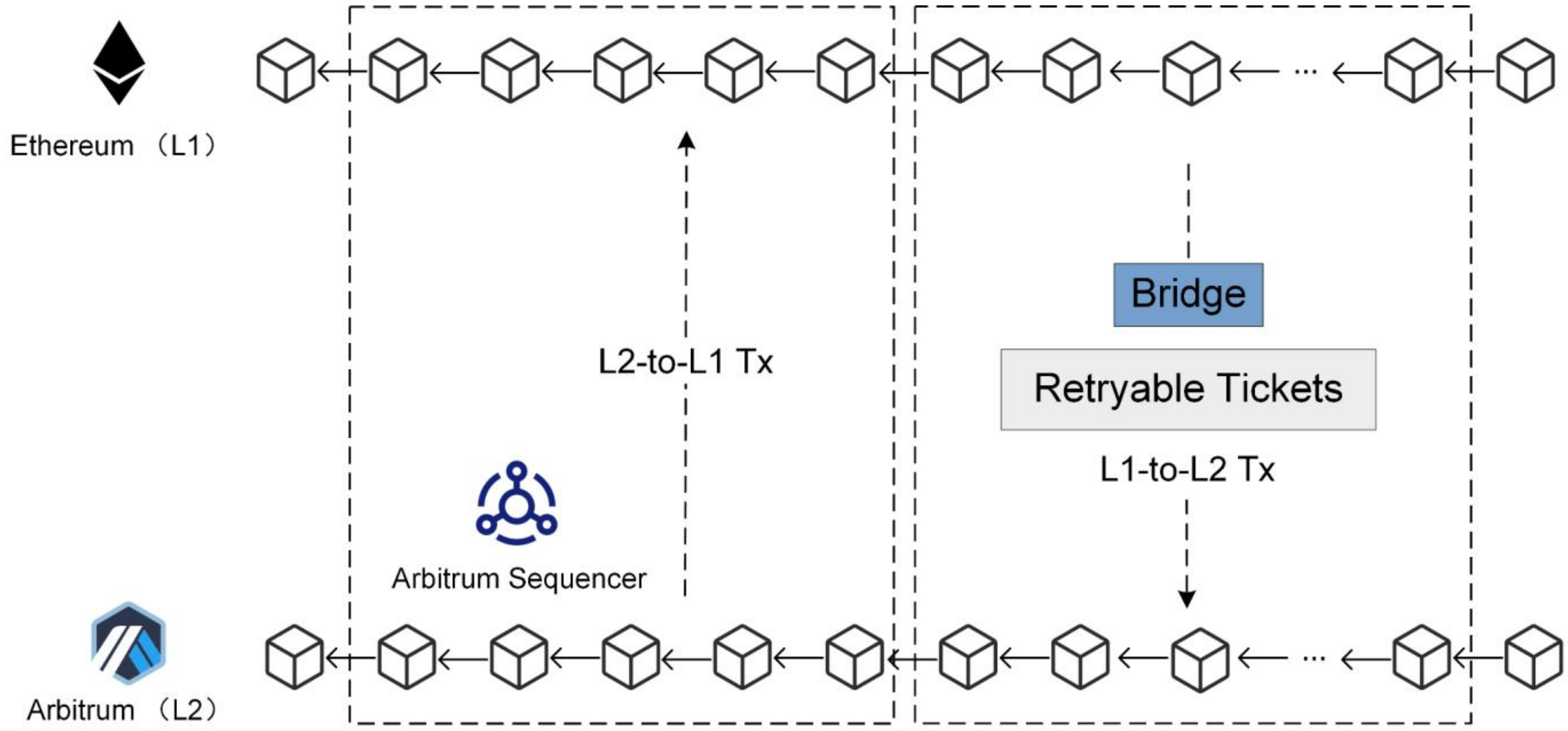

**Figure 3.** Transaction Classification Diagram. L2-to-L1 transactions require the involvement of the Arbitrum sequencer, while L1-to-L2 transactions are implemented through the bridge and retryable tickets.

In the following content, we will delve into the details of these transaction types and the potential security risks that may arise.

***Sequencer in L2-to-L1 Message***

The sequencer is a specifically designated full node in Arbitrum. Its main responsibility is to collect, order, and execute users' L2 transactions, providing immediate transaction results and receipts to users. Additionally, the sequencer periodically bundles multiple transactions and submits them to Ethereum to enhance the efficiency of the entire system. Regarding the Sequencer, the issue we discuss is the impact of its status on transaction execution.

As shown in Figure 4, when the sequencer experiences downtime, transactions will accumulate in the delayed inbox and cannot be executed immediately until the sequencer is restored. Once the sequencer is back online, it prioritizes processing all the old transactions in the delayed inbox before handling new transactions. However, these old transactions will still experience delayed execution.

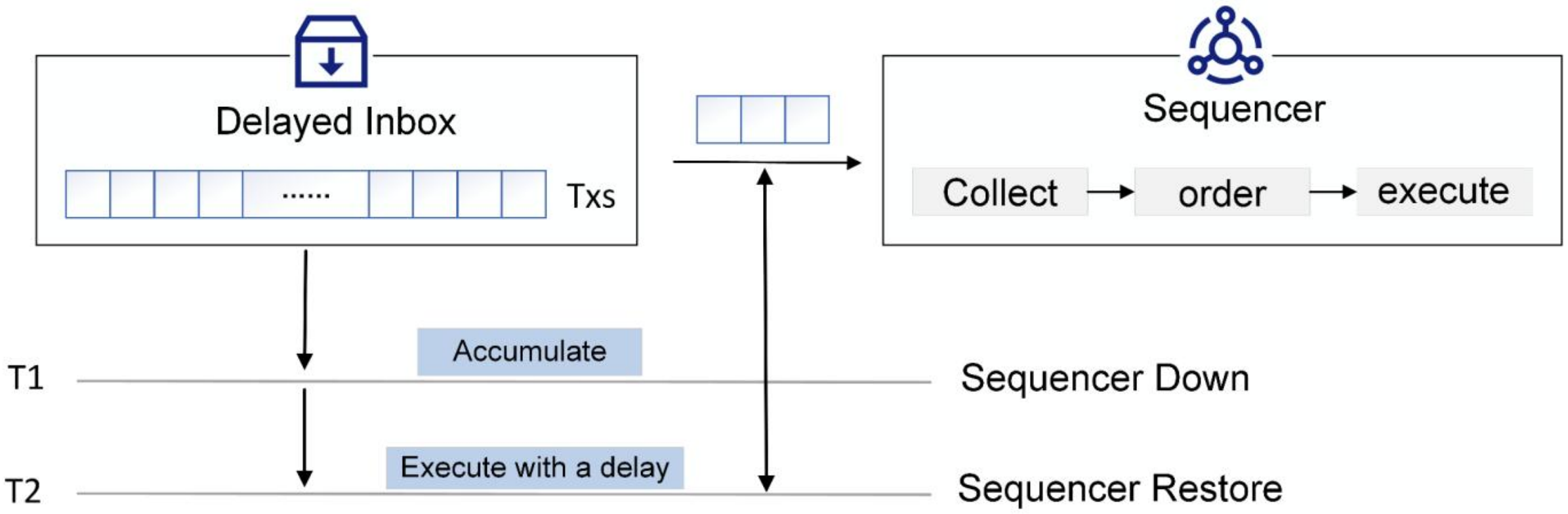

**Figure 4.** Transaction Delay Execution Diagram. The sequencer's downtime results in transaction accumulation and delayed execution, compromising the real-time nature of transaction execution.

In addition to sending L2 transactions to the sequencer, users or projects can also directly send L2 transactions to Ethereum's delayed inbox. The delayed inbox is a smart contract deployed on Ethereum mainnet, protected by Ethereum's smart contract security mechanisms. However, it is worth noting that transactions sent directly to the delayed inbox by users will also experience delays until a specified time or block height, when they ultimately need to be executed by the sequencer. Thus, the execution time of these transactions will be delayed. Furthermore, when sending transactions to the delayed inbox, users need to manually set transaction parameters such as gas price and gas limit to ensure timely execution. Improper parameter settings may lead to transaction execution failures or the risk of overpaying transaction fees.

Issues and Risks: Considering that L2-to-L1 transactions rely on the sequencer to process transactions and bundle blocks, when migrating smart contract code from Ethereum to Arbitrum, it is crucial to consider the operational status of the sequencer, especially when the code involves obtaining real-time off-chain data. If the sequencer experiences downtime and reconnects, it may result in various issues, such as:

1. Incorrect transaction execution: If a contract relies on real-time off-chain data for executing transfers or other fund-related operations, delayed data can cause transaction execution delays or errors. This can lead to funds being transferred incorrectly or transactions not being executed as expected.

2. Asymmetric transaction conditions: Delayed off-chain data may cause information asymmetry between smart contracts and external data sources, potentially leading to missed opportunities or unfair transaction conditions. This can result in certain participants obtaining unfair advantages, leading to financial losses for other participants.

3. Inaccurate prices or market data: If a smart contract relies on timely prices or market data for executing trades or making decisions, delayed or inaccurate data can lead to transactions being executed at inappropriate prices or market conditions, resulting in financial losses.

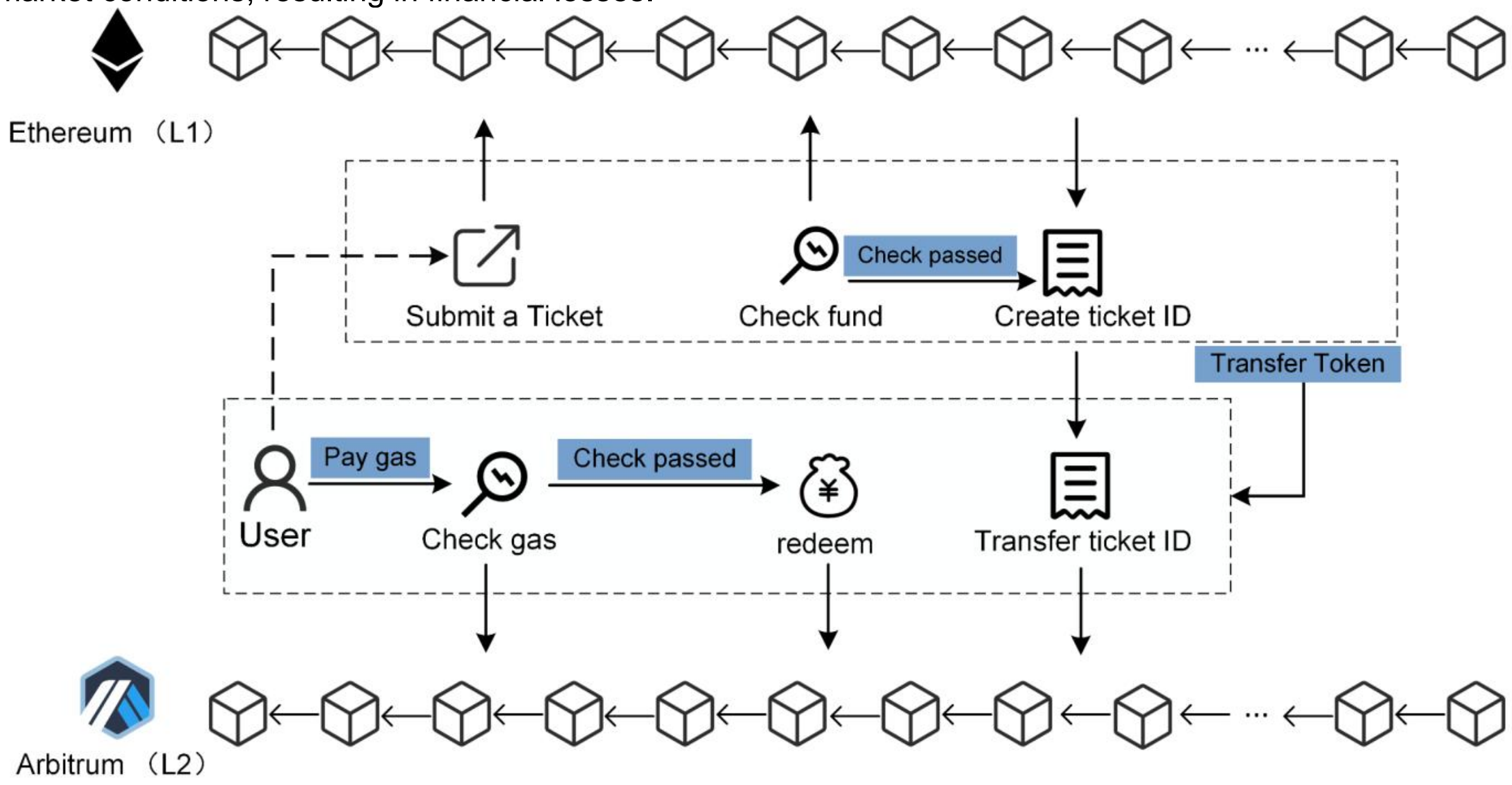

**Figure 5.** Illustration of L1-to-L2 Messaging Invocation. Asset transfer is achieved based on ticket generation on L1 and the redeem operation on L2.

***L1-to-L2 Messaging through Retryable Tickets***

The invocation of L1-to-L2 messaging can be done through retryable tickets, which is generally used for asset transfer. This process can be divided into two stages: submitting a ticket on L1 (calling the Inbox.createRetryableTicket() function); and executing the redeem operation on L2. The submission on L1 and the execution on L2 are asynchronous and can be separated.

1. Submission process on L1.

Submit a ticket on L1, which will generate a unique ticket ID. This process is asynchronous, meaning that users can continue with other operations after submitting the ticket without waiting for the execution on L2. Sufficient funds need to be provided for the ticket execution fees, and a fund check is also required. If the check passes, it indicates that the ticket creation was successful.

2. Redeem process on L2.

After successfully creating the ticket, the redeem operation is performed on L2, transferring the assets from L1 to L2. Before execution, a gas check is required. If the check passes, the redeem operation is automatically executed. If the automatic redemption fails, the redeem operation needs to be performed manually. Figure 5 illustrates the L1-to-L2 messaging invocation.

The redeem operation is automatically executed by the smart contract on L2. In this process, there are two possible scenarios. If the automatic redemption is successful, the user's specified L2 address receives a refund of the submission fee. However, if the redemption fails, the submission fee is charged, and only the remaining gas fee after execution is returned to the L2 address specified by the user. Additionally, if automatic redemption fails, the ticket is temporarily stored in the L2 buffer for one week, and the submission fee is associated with this temporary storage. If automatic redemption is successful, it will not be stored in the buffer, and therefore, no submission fee needs to be paid.

Issues and Risks: In the L1-to-L2 messaging transfer, there is a possibility of failure during the submission process on L1 or the redeem process on L2, which may lead to adverse consequences:

1. When calling the createRetryableTicket() function on L1, if the funds fail the verification check, the transaction will be reverted, resulting in the loss of gas fees without any refund.

2. After the automatic redemption fails on L2, a manual redeem operation is required. This may cause:

   Overpayment of submission fees: In the case of automatic redemption success, the submission fee is refunded. However, in the case of manual redemption, the submission fee needs to be paid again, resulting in multiple fee payments.

   Delay in transaction execution time: Manual redeem operation after automatic redeem failure may cause a delay in the execution time of the transaction, which could lead to losses such as price fluctuations.

   Payment of L2 gas fees: Manual redeem operation also requires the payment of L2 gas fees to ensure proper execution of the transaction on L2.

3. If the ticket fails to redeem automatically and is temporarily stored in the buffer, it will be saved for seven days. If the ticket fails to redeem or the fee is not paid to continue the storage after the expiration, the carried messages and value (excluding the managed callvalue) may be lost and cannot be recovered.

**Block Properties**

Block properties refer to specific attributes or metadata that each block in Arbitrum possesses. These properties provide information about the block itself and the transactions contained within it. Examples of block properties include block height, timestamp, hash value, and the number of transactions included.

In Arbitrum and Ethereum, block properties generally refer to the block summary data stored in the block header. However, there may be some differences in the results obtained when retrieving Ethereum block properties. Let’s take the block number and the block.timestamp as examples.

Table 2 shows the results obtained from Ethereum and Arbitrum when retrieving the Ethereum block number based on block.number. The Ethereum block time interval is approximately 15 seconds, while the Arbitrum sequencer requests an update of the L1 block number every minute. This means that the L1 block number obtained by Arbitrum will remain unchanged during this one-minute interval. In the 12:00-12:01 am interval, the block number obtained is consistently 1000.

Results obtained from L1 Ethereum and L2 Arbitrum based on block.timestamp also differ. On Ethereum, the returned timestamp represents the current block’s timestamp, while on Arbitrum, the timestamp recorded by the sequencer is obtained.

| Wall Clock time | L1 block.number | L2 block.number |
|---|---|---|
| 12:00:00am | 1000 | 1000 |
| 12:00:15am | 1001 | 1000 |
| 12:00:30am | 1002 | 1000 |
| 12:00:45am | 1003 | 1000 |
| 12:01:00am | 1004 | 1004 |
| 12:01:15am | 1005 | 1005 |

**Table 2.** Comparison of Wall Clock Times, L1–block.number, and L2–block.number. Data Source: https://docs.arbitrum.io/time

Below, we analyzed the reasons for the aforementioned differences. As shown in Figure 6, when the sequencer packs transactions, it gathers various data. Three aspects of the data are relevant to our research: the current L1 block number obtained from Ethereum, which represents the latest block number on Ethereum; the timestamp recorded locally by the sequencer, which is used to assess the synchronization status of Arbitrum by comparing it to the l1 Timestamp obtained from Ethereum; and a certain number of transactions collected by the sequencer. If there are no issues, the sequencer will pack the L1 Block Number, timestamp, and these transactions together into a block.

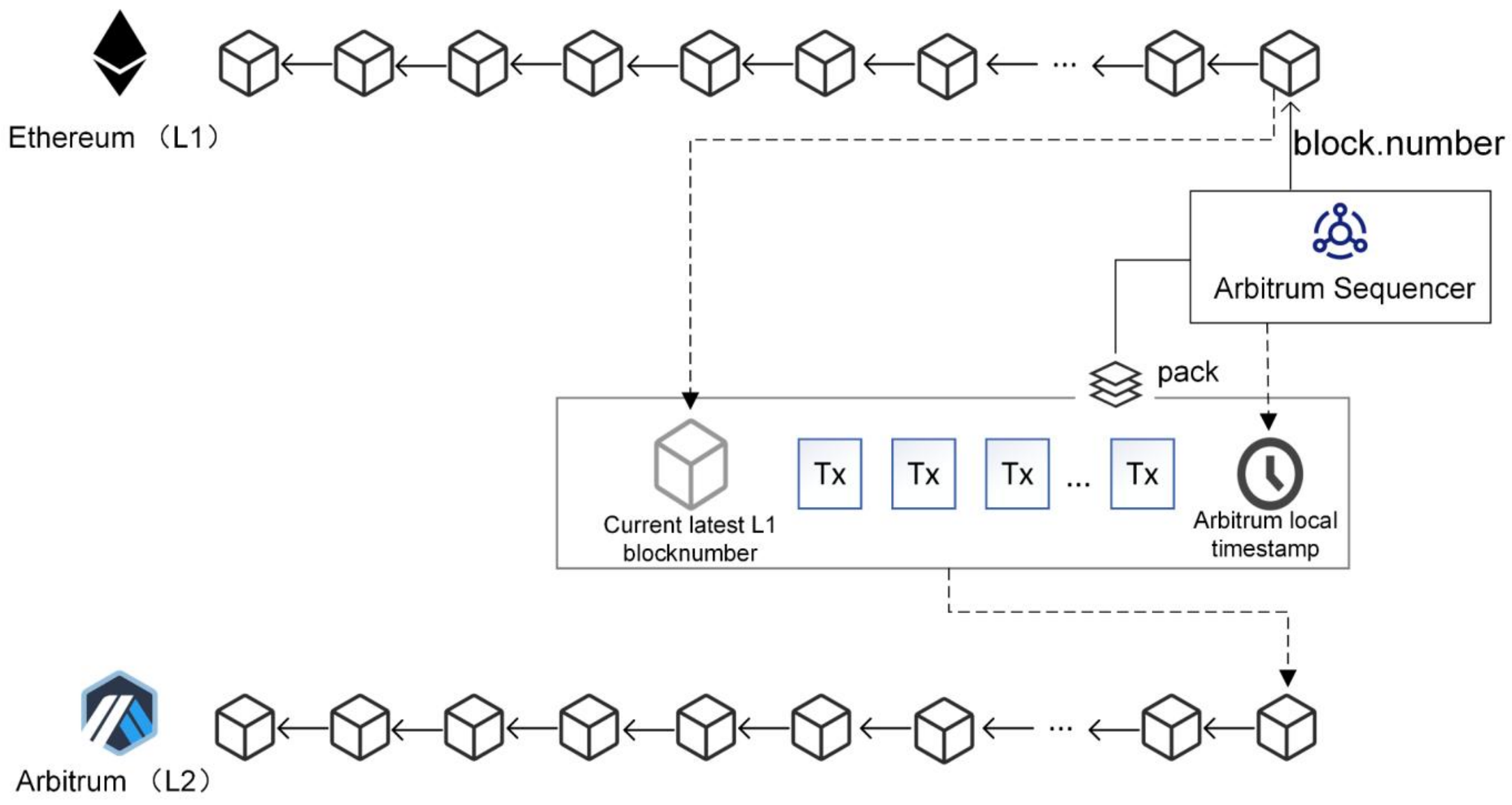

**Figure 6.** Block Packaging Illustration. Among the data dependencies for block packaging, three aspects are relevant to our research: L1 Block Number, Local Timestamp, and Txs (transactions).

The method used by the sequencer to retrieve the local timestamp may return timestamps with low precision, and repeated retrieval within a very short time may result in the same timestamp being returned. In Arbitrum, the sequencer obtains the local timestamp to determine the block timestamp. Unlike Ethereum, Arbitrum does not impose a time difference between new blocks and their parent blocks. If the sequencer reads the time too frequently, it may result in different blocks having the same timestamp on Arbitrum.

Generally speaking, the acquisition of block numbers by smart contracts on Ethereum is continuous, while on Arbitrum (acquiring Ethereum's block numbers), it is discontinuous, with the value updated every minute. Because of this, if a smart contract on the Arbitrum uses a strict equality check (for example, currentNumber==customNumber), problems may arise. This is due to the fact that a specific customNumber value might not exist on the Arbitrum, so there's a certain probability that the conditions for the strict equality check cannot be met.

Issues and Risks: On Ethereum and Arbitrum, there are certain differences in the block properties obtained based on block.number and block.timestamp. These differences can lead to inconsistent behavior of smart contracts on Arbitrum and Ethereum, which may result in incorrect outcomes, security vulnerabilities, or other unpredictable behavior. Therefore, careful consideration of these differences and their potential security implications is necessary when migrating smart contracts from Ethereum to Arbitrum.

1. Inconsistent block number logic: Due to the different mechanism for updating L1 block numbers in Arbitrum compared to Ethereum, smart contracts may not be able to make accurate logical judgments based on block numbers. For example, if a contract relies on the state or events of a specific block, inconsistent results may occur on Arbitrum.

2. Inconsistent timestamp logic: Since the timestamp obtained in Arbitrum is recorded by the sequencer and not the timestamp of the current block, smart contracts may not be able to make accurate logical judgments based on timestamp. For example, if a contract needs to perform certain operations or restrictions based on the current time, inconsistent results may occur on Arbitrum.

3. Duplicate timestamp issue: It is possible for Arbitrum to have the same timestamp in different blocks, which can prevent contracts from accurately differentiating between timestamps of different blocks. This can impact time-related contract logic, such as restricting certain operations to be executed only once within a specific time period.

**Contract Address Alias**
Smart contract address alias refers to the new address assigned to a contract after migrating from Ethereum to Arbitrum. Setting an alias address for a smart contract is done to mitigate the risk of contract impersonation and ensure the security of cross-chain information transmission and interaction between Ethereum and Arbitrum.

When a smart contract is migrated from Ethereum to Arbitrum, it is assigned a new address, which results in a difference in ‘msg.sender‘. The ‘msg.sender‘ is commonly used in smart contract code to return the address of the message sender. As shown in Figure 7, in L2-to-L2 transactions on Arbitrum, ‘msg.sender‘ will return the address of the current message sender, consistent with the behavior on Ethereum. However, in L1-to-L2 transactions, ‘msg.sender‘ will return the address alias of the L1 contract that triggered the message within L2.

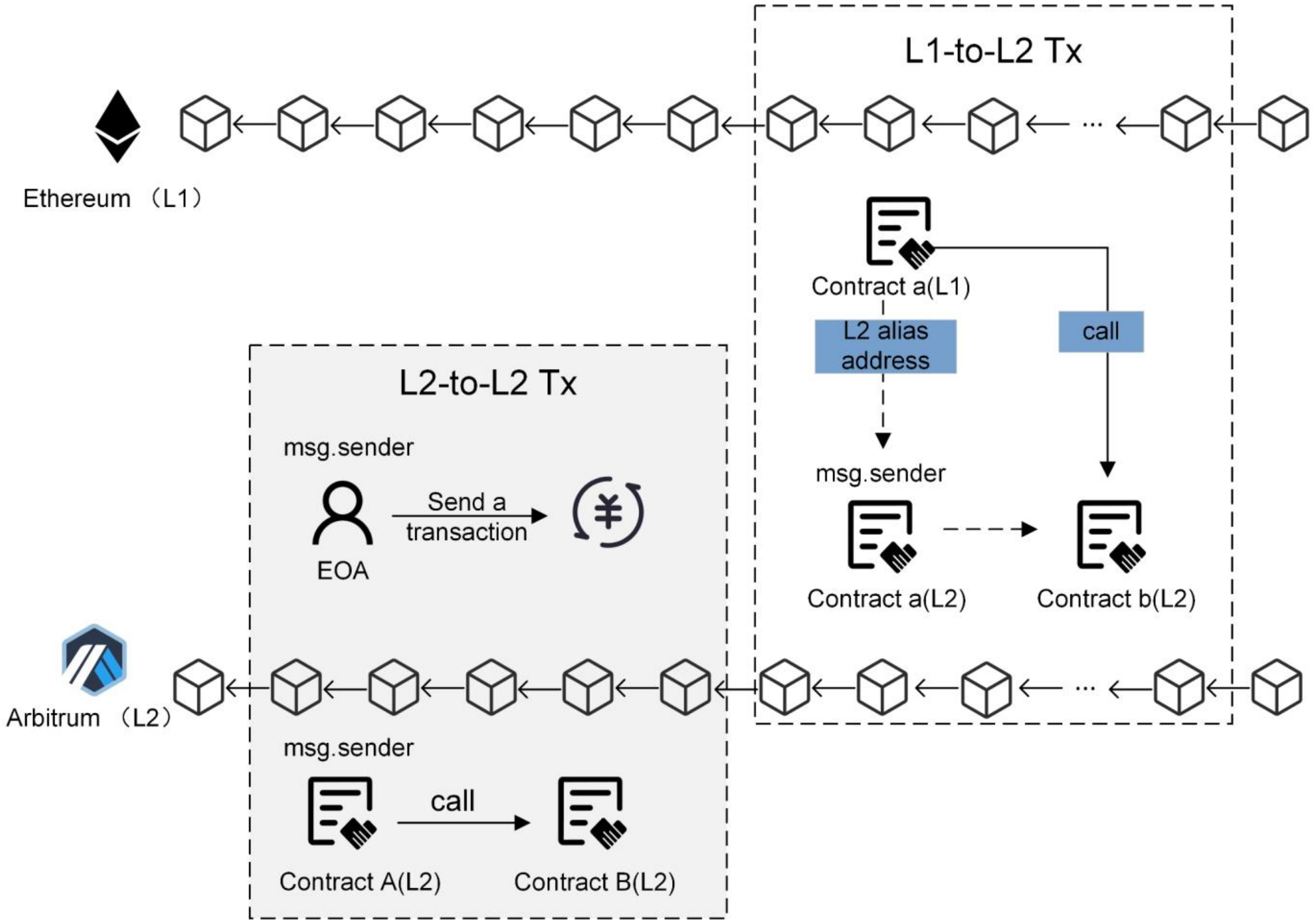

**Figure 7.** Diagram of Retrieving ‘msg.sender‘. The ‘msg.sender‘ returns the address of the message sender. However, it is important to note that on Arbitrum, the message sender may have an alias address.

The smart contract alias address calculation method is as follows: A represents the contract’s address on L1, C is a constant, and f(A) represents the alias address of the contract on L2.

$$f(A) = (A+C) \bmod 2^{160}$$

$$C = 0x1111000000000000000000000000000000001111$$

The reason for setting an alias for the L1 contract address on Arbitrum is to avoid contract impersonation risks. Specifically, when a contract on Ethereum submits a transaction to Arbitrum, a problem arises regarding which sender address should be attached when the contract is running on Arbitrum. The simple solution is to use the L1 address of the sending contract, but there may be another contract on Arbitrum with the same address. If such a contract exists, the calling receiver on Arbitrum will not be able to distinguish between these two contracts, allowing one contract to impersonate another, which poses potential risks.

Issues and Risks: When migrating smart contracts from Ethereum to Arbitrum, it is important to be aware that if the contract involves L1-to-L2 message passing or comparing constant addresses with ’msg.sender’, the alias address obtained from L1-to-L2 should be considered. If developers overlook the fact that the L1-to-L2 message retrieves an alias address, it may lead to the following security risks:

1. Failed permission checks: If a contract on Arbitrum performs permission checks by comparing 'msg.sender' with an expected address and does not take into account the alias address from L1-to-L2 messages, permission checks may fail. This means that unauthorized addresses may be granted access to the contract, resulting in potential security vulnerabilities.

2. Inability to modify contract owner: In some cases, a contract may need to modify its owner address. However, if the contract's owner address is obtained through L1-to-L2 messages and developers do not handle the alias correctly, the contract may be unable to modify its owner. This can limit the contract's functionality and flexibility, as it may not be able to update its logic or configuration.

3. Potential contract impersonation risks: If a contract on Arbitrum uses 'msg.sender' for contract impersonation risk checks and does not consider the alias address, there may be a risk of contract impersonation. This means that one contract can impersonate another and perform unauthorized operations, potentially resulting in financial losses or other adverse effects.

**Gas Fees**

Gas fees refer to the resource consumption required to execute smart contracts on blockchain platforms such as Ethereum and Arbitrum. Gas fees are necessary to compensate validators for their computational work and to prevent abuse or inefficient code execution. Each gas unit has a corresponding gas price, and miners determine the priority of transaction inclusion based on the gas price.

Gas fees consumed when executing smart contracts on Arbitrum is lower compared to Ethereum. This is because Arbitrum transactions are executed in batches, where multiple transactions are combined into a single batch for processing, reducing the fees and gas consumption per transaction. Additionally, Arbitrum moves a portion of the computation process off the Ethereum and processes it in an off-chain environment, further reducing the computational load and gas consumption on Ethereum.

We have obtained gas fees data for some transactions on Ethereum and Arbitrum, as shown in Table 3. For example, if a user performs a deposit operation on the Aave protocol deployed on Ethereum, it would cost $4.02. However, if the same operation is performed on the Aave protocol deployed on Arbitrum One, it would only cost $0.15, resulting in a 96% cost savings. The average gas consumption for the six transactions in the table on Arbitrum one is 95.5% lower than Ethereum.

| Tx Example | Arbitrum One | Ethereum | Pct saved | Amount saved |
|---|---|---|---|---|
| Aave Deposit | $ 0.15 | $ 4.02 | 96% | $3.87 |
| EOA Transfer | $ 0.09 | $ 0.65 | 87% | $0.57 |
| Opensea NFT Sale | $ 0.20 | $ 5.55 | 96% | $5.35 |
| SushiSwap Swap | $ 0.08 | $ 2.53 | 97% | $2.45 |
| Uniswap Swap | $ 0.08 | $ 3.97 | 98% | $3.89 |
| Yearn Deposit | $ 0.05 | $ 3.63 | 99% | $3.58 |

**Table 3.** Gas fees Comparison between Arbitrum and Ethereum (Pct stands for Percentage). Data Source: https://gas.arbitrum.io/

Gas fees on Arbitrum consist of two aspects: firstly, the calldata of transaction invocation on Ethereum, which is intended to compensate the sequencer for the cost of publishing transactions on Ethereum, and secondly, the

Arbitrum portion assessed and collected by ArbOS. The gas calculation formula is as follows:

$$gasLimit = gasUsed_{L2} + (calldataPrice_{L1} * calldataSize_{L1})/gasPrice_{L2}$$

$$gasFees = gasLimit * gasPrice_{L2}$$

In the formula, $gasUsed_{L2}$ stands for the amount of gas used for executing transactions on Arbitrum L2 platform. $calldataPrice_{L1}$ stands for the price of L1 calldata, indicating the cost of invoking transactions on Ethereum L1 platform. $calldataSize_{L1}$ stands for the size of L1 calldata, indicating the number of bytes of data passed to the Arbitrum L2 sequencer. $gasPrice_{L2}$ stands for the price of gas used on Arbitrum L2 platform. The calculated *Gaslimit* represents the total amount of gas used for transactions on Arbitrum L2 platform. The calculated *gasFees* represent the gas fees required for executing transactions on Arbitrum L2 platform.

Gas fees on Arbitrum are not fixed and can vary. This is because each block on Arbitrum has a fixed block space and gas limit, while the number of transactions within a block is uncertain. When there are more transactions within a block, the gas allocated per transaction decreases, and vice versa.

Gas fees required for transactions on Ethereum are relatively high, making it less likely for attackers to successfully launch a large number of small-value transactions to attack Ethereum. However, due to the lower gas fees on Arbitrum sometimes, attackers can carry out attacks involving a large number of small transactions. Therefore, it is still important to remain vigilant to ensure the security of our assets.

Issues and Risks: When gas fees on Arbitrum are relatively low, it can lead to DOS attacks.

1. Large-scale small-value transaction attacks: Due to the lower gas fees, attackers may exploit the Arbitrum to launch attacks involving a large number of small-value transactions. Such attacks can cause network congestion, transaction delays, and resource waste, negatively impacting the network's normal operation and user experience.
2. Resource exhaustion: Large-scale small-value transaction attacks can deplete network resources, including computational and storage resources. This can result in increased transaction processing times, transaction failures, or DOS issues, affecting users' normal transaction activities.
3. Exploitation of malicious contracts: Attackers may create malicious smart contracts to exploit the low gas fees and execute a large number of transactions for certain purposes, such as market manipulation, fraudulent activities, or other improper behaviors. This can cause significant losses and inconvenience to users and the ecosystem.

### Summary of Issues

Based on the characteristics of Arbitrum and the differences between Arbitrum and Ethereum, we have summarized the issues that may arise when migrating smart contracts from Ethereum to Arbitrum, depending on what logic the smart contract contains.

#### *Outdated off-chain data obtained by the inactive sequencer:*

Smart contracts that include logic to fetch off-chain real-time data and rely on the proper functioning of the sequencer need to be migrated from Ethereum to Arbitrum. If the sequencer is in an abnormal state and fetches outdated data, this can lead to incorrect transaction execution, asymmetric trading conditions, and inaccurate price or market data, resulting in financial losses and unfair trading conditions.

#### *Logic errors based on time:*

Smart contracts containing the logic to retrieve block number and timestamp are being migrated from Ethereum to Arbitrum. This logic has resulted in inconsistencies in the behavior of the contract between Ethereum and Arbitrum, causing issues when dealing with time-related operations.

#### *The permission check failed:*

On Arbitrum, ignoring address aliasing in L1-to-L2 transactions can lead to security issues during the migration process of smart contracts that involve permission checks logic. These issues may include permission check failures, inability to modify contract owners, and potential contract impersonation risks.

#### *DOS attack:*

When gas fees are low on Arbitrum, there may be security risks such as large-scale, small-value transaction attacks, resource exhaustion, and exploitation of malicious contracts.

To validate the aforementioned research findings, we conducted a case study analysis. The specific details can be found in the supplementary information.

## Risk Avoidance Methods

**Outdated off-chain data obtained by the inactive sequencer:**
When a smart contract running on Arbitrum needs to obtain real-time data, to avoid the issue of obtaining outdated data due to a Sequencer outage, Chainlink L2 Sequencer Uptime Feeds can be used. This involves checking whether the Sequencer is in normal operation before relying on it to fetch data from off-chain. Only a Sequencer that is functioning normally can acquire timely off-chain data.

**Logic errors based on time:**
To prevent unexpected errors in operations on Arbitrum that depend on block information like block numbers and timestamps, smart contracts running on Arbitrum should be mindful of potential delays when using block.number. Also, when using block.timestamp, it is important to be aware that the accuracy of the timestamps might not be high, which can lead to data inconsistencies.

In light of these issues, it's crucial to avoid using block information such as block.number and block.timestamp in scenarios where time sensitivity is critical. Furthermore, it's essential not to hardcode block numbers in smart contract code, which means not using specific block numbers for performing operations or querying specific blocks on the blockchain.

**The permission check failed:**
If a smart contract's permission checks are based on the contract's address, it's crucial to pay special attention to the characteristic of address aliasing. Overlooking this aspect can lead to errors in permission checks. When migrating a smart contract from Ethereum to Arbitrum, it is necessary to set an alias for the smart contract. It's important to note that when msg.sender is retrieved, the returned address is the alias address of the contract on Arbitrum, not its address on Ethereum.

**DOS attack:**
To reduce the likelihood of DoS attacks, for certain sensitive operations or users participating in a contract, introducing a fee or requiring a certain amount of collateral can be effective. Attackers would need to pay this fee to execute operations, thus increasing the cost of the attack.

In collateralized lending contracts, increasing the minimum deposit amount and introducing deposit time intervals can prevent frequent small deposits. Also, introducing a deposit threshold that requires the deposit amount to reach a certain level before updating the collateral ratio can be beneficial.

In crowdfunding projects, auction projects, or other contracts where batch refunds or other batch operations might occur, there may be arrays without explicit size limitations or constraints. Attackers could target a function or contract that processes arrays by adding a large amount of data to the array, leading to gas depletion during the loop process and causing the function to fail. To counter this, limiting the size of arrays can prevent attackers from adding excessive data that leads to too many iterations. Alternatively, using mappings instead of arrays and iterating over the keys of the mapping for operations can avoid the need for loops.

## Limitations

**Data Collection:**
We collected data including relevant literature, official online resources, and smart contract source code. We analyzed 6 typical cases of vulnerabilities or risks found in the collected data based on the underlying differences between Ethereum and Arbitrum. We also summarized the risks of migrating smart contracts from Ethereum to Arbitrum. However, due to limited data on smart contract migration, there may still be some potential risks that have not been disclosed. Additionally, it is important to note that there are no smart contracts that are absolutely free of vulnerabilities. We must be aware that during the migration process, we may encounter unknown risks.

**Selection of Research Objects:**
Our research focuses on Ethereum and Arbitrum. In the current blockchain technology landscape, there are various L1 and L2. L1 includes platforms like Hyperledger Fabric and EOS, while L2 includes solutions like Optimism and Polygon. The demand for smart contract migration exists within these L1 and L2. Different migration paths (i.e., from source blockchain to target blockchain) have underlying differences that introduce different security risks. We did not cover all possible migration paths.

It is worth noting that we were fortunate to receive support from the Arbitrum Office Lab, which provided us with rich and authentic reference materials. These materials were crucial for our research, as they helped us conduct a more comprehensive and reliable research.

**Migration Risks:**
The smart contract vulnerabilities caused by migration originate from the contract's execution environment. These vulnerabilities do not exist before migration. Once the contract is migrated to another blockchain, the security risks appear. This is due to the underlying differences between the source and target blockchains, resulting in a change in the execution environment of the smart contract. Therefore, to effectively mitigate security issues caused by smart contract migration, auditors need to have specialized knowledge. Auditors should carefully analyze the blockchain protocols, virtual machines, and other related components of both the source and target blockchains. They need to understand the differences between different blockchains, including implementation details of the underlying protocols, features and limitations of smart contract languages, and behavioral differences of virtual machines, among others. Only with a deep understanding of these differences can auditors accurately assess the potential security risks during the contract migration process and provide relevant recommendations and solutions.

## Conclusions and Future Work

There are security risks involved in migrating smart contracts deployed on Ethereum to Arbitrum, which we have studied. In this research, we analyzed various security issues related to smart contract migration, including the features of Arbitrum, the differences between Arbitrum and Ethereum, and multiple vulnerable smart contract cases such as outdated off-chain data obtained by the inactive sequencer, logic errors based on time, the permission check failed, DOS attack. Furthermore, we provided we proposed avoidance methods and provided considerations for these potential security risks.

This paper primarily focuses on security risks during the migration process of smart contracts from Ethereum to Arbitrum. However, we must also acknowledge that apart from the smart contract migration issues discussed in this paper, there are other security risks. In fact, the security of smart contract migration is an evolving field with new challenges and potential risks constantly emerging. The security issues in smart contract migration may present even more challenges and risks when migrating between different blockchains. Whether it is migrating from Ethereum to Arbitrum, Ethereum to other Ethereum L2 scaling solutions (such as Optimism or Starknet), or even between Ethereum and other homomorphic chains (like Binance Smart Chain or Polygon), there are potential security risks involved.

Therefore, achieving secure smart contract migration requires continuous effort and exploration. The security issues in smart contract migration need to be continuously monitored, researched, and addressed. We recommend users carefully evaluate the characteristics, security, and feasibility of the migration target blockchain, as well as the underlying differences between the source and target blockchains, before proceeding with smart contract migration. It is also advisable to engage in discussions and consultations with relevant technical experts and communities to ensure a smooth migration process that aligns with the project's requirements.

## Acknowledgements

Thank you to Mehdi Salehi and ChrisCo from Offchain Labs for providing advice, reference materials, and conference support.